\documentclass[aps,prd,onecolumn,groupedaddress,showpacs,nofootinbib,amssymb]{revtex4}
\usepackage{graphicx,bm,color}
\usepackage{amsmath}
\usepackage{amssymb}
\usepackage{amsfonts}
\usepackage{epsfig}
\newcommand{\be}{\begin{equation}}
\newcommand{\ee}{\end{equation}}
\newcommand{\bea}{\begin{eqnarray}}
\newcommand{\eea}{\end{eqnarray}}
\newcommand{\beaa}{\begin{eqnarray*}}
\newcommand{\eeaa}{\end{eqnarray*}}

\newcommand{\nn}{\nonumber \\}

\allowdisplaybreaks[4]

\begin{document}

\title{Cosmology and stability in scalar tensor bigravity \\with non-minimal kinetic coupling gravity}
\author{F. Darabi}\email{ f.darabi@azaruniv.edu}
\author{ M. Mousavi}\email{ mousavi@azaruniv.edu}

\affiliation{Department of Physics, Azarbaijan Shahid Madani University, Tabriz, 53714-161 Iran}

\begin{abstract}

We generalize the scalar tensor bigravity models to the non-minimal kinetic coupling scalar tensor bigravity models with two scalar fields whose kinetic terms are non-minimally coupled to two Einstein tensors constructed by two metrics. We show that a broad class of expanding universes can be explained by some solutions of this model. Then, we study the stability issue of the solutions by means of imposing homogeneous perturbation on the equations of motion and extract the stable solutions.

\end{abstract}


\maketitle

\section{Introduction \label{Sec1}}

The infrared modifications of Einstein general relativity is receiving attention by two aspects: i) the theoretical aspect from an effective field theory point of view that is one of the very natural choices to seek when declaring the diffeomorphism invariance in GR, and ii) the observational aspect by means of explaining the accelerating expansion of the Universe. In this regard, it seems that general relativity has to include a new mass term to validate such attempts. Recently, a ghost free \cite{1,2} non linear massive gravity model (the dRGT theory) was constructed \cite{3,4,5,6,7}  involving two metric tensors, one dynamical $g_{\mu\nu}$ and one non-dynamical $f_{\mu\nu}$ appearing in a unique set of terms. These two metrics are used to add a new mass term to the original action (GR), namely the interaction term. For this model with a non-dynamical reference metric, it has been shown that the flat Friedmann-Robertson-Walker
(FRW) Universe dose not exist \cite{8}, and that the open FRW solutions are allowed \cite{9} involving the problems of strong coupling \cite{10} and ghostlike instabilities \cite{11}. Attempts to screen such problems led people to a natural way of extending the massive gravity theory and going beyond it toward a similar new model in which  two dynamical symmetric tensors $g_{\mu\nu}$ and  $f_{\mu\nu}$ appear as foreground and background metrics, respectively in a completely symmetric manner \cite{12,13,14,15} which is called the bigravity theory. Obviously, the massive bigravity theory covers the massive gravity theory and treats with two metrics thoroughly in a symmetric way such that convinces one to get rid of the aether-like concept of reference metric in massive gravity. Cosmology in massive gravity model has been studied in Refs. \cite{16,17,18,19}, whereas in bigravity several branches of regular cosmological solutions have been extracted \cite{20,21,22'}. Afterwards, the bigravity model and its modification with two independent scalar fields have been investigated in \cite{22,23,23'} presenting the models which certify the stable solution describing the spatially flat FRW solution. In these models, the scalar fields are minimally coupled to the metric. As we know, there are varieties of cosmological models which contain scalar fields non-minimally coupled to gravity \cite{24,25,26}. Moreover, one can go on the extension of scalar-tensor theories and construct coupling between the derivatives of the scalar fields and the curvature, namely the non-minimal kinetic coupling scalar tensor theory \cite{27}. In this work, we consider such non-minimal kinetic coupling scalar tensor bigravity model as a generalization of the model studied in \cite{22}.

The organization of this paper is as follows. In section 2, we review the field equations of motions for the minimal bigravity action. In section 3, we construct bigravity models with two scalar fields forming the action of non-minimal kinetic coupling scalar tensor bigravity model and again obtain the equations of motions besides extracting the Bianchi constraints. In section 4, we show that a broad class of the expansion history of the universe can be explained by some solutions of the bigravity model, whereas these solutions do not have always stability against the  perturbation. In section 5, we go through the stability issue of the solutions. The paper ends with a conclusion.

\section{Bigravity Theory \label{Sec2}}

The action of Hassan-Rosen theory named bigravity has the following structure
\cite{28}
\be
\label{Fbi1}
S_{{\rm bi}}=M_{g}^{2}\int d^{4}x\sqrt{- {\rm det} g}R^{(g)}+M_{f}^{2}\int d^{4}x\sqrt{- {\rm det} f}R^{(f)}+
2m^{2}M_{{\rm eff}}^{2}\int d^{4}x\sqrt{- {\rm det} g}\sum_{n=0}^{4}\beta_{n}e_{n}\left(\sqrt{g^{-1}f}\right).
\ee
Here $g_{\mu\nu}$ and $f_{\mu\nu}$ are two dynamical tensors in the gravity sector, and  $R^{(g)}$ and $R^{(f)}$ are the scalar curvatures corresponding
to the metric tensors $g_{\mu\nu}$ and $f_{\mu\nu}$, respectively. It should be noted that $M_{{\rm eff}}$ is defined as

\begin{align}
\label{Fbi2}
\frac{1}{M_{{\rm eff}}^{2}}=\frac{1}{M_{g}^{2}}+\frac{1}{M_{f}^{2}} .
\end{align}
The tensor $\sqrt{g^{-1}f}$ is the square root of $g^{\mu\rho}f_{\rho\nu}$ which means, $\left(\sqrt{g^{-1}f}\right)^{\mu}~_{\rho}\left(\sqrt{g^{-1}f}\right)^{\rho}~_{\nu}=g^{\mu\rho}f_{\rho\nu}=X^{\mu}~_{\nu}$. For this defined tensor, $e_{n}(X)$'s are given by
\begin{align}
\label{Fbi3}
e_{0}(X)=&1,~~e_{1}(X)=[X],~~e_{2}(X)=\frac{1}{2}\left([X]^{2}-[X^{2}]\right),\nn
e_{3}(X)=&\frac{1}{6}\left([X]^{3}-3[X][X^{2}]+2[X^{3}]\right),\nn
e_{4}(X)=&\frac{1}{24}\left([X]^{4}-6[X]^{2}[X^{2}]+3[X^{2}]^{2}+8[X][X^{3}]-6[X^{4}]\right),\nn
e_{k}(X)=&0~~{\rm for}~~k>4,
\end{align}
where $[X]$ means the trace of the tensor $X^{\mu}~_{\nu}$.
For simplicity, we take the minimal and non-trivial case as \cite{29}
\begin{align}
\label{Fbi4}
S_{{\rm bi}}=M_{g}^{2}\int d^{4}x\sqrt{- {\rm det} g}R^{(g)}+M_{f}^{2}\int d^{4}x\sqrt{- {\rm det} f}R^{(f)}+2m^{2}M_{{\rm eff}}^{2}\int d^{4}x\sqrt{- {\rm det} g}\left(3-{\rm tr}\sqrt{g^{-1}f}+{\rm det }\sqrt{g^{-1}f}\right),
\end{align}
which is apparently a direct result of using equation (\ref{Fbi3}) in terms of $e_{n}$ as follows
\begin{equation}
3-{\rm tr}\sqrt{g^{-1}f}+{\rm det }\sqrt{g^{-1}f}=3e_{0}\left(\left(\sqrt{g^{-1}f}\right)^{\mu}~_{\nu}\right)-e_{1}\left(\left(\sqrt{g^{-1}f}\right)^{\mu}~_{\nu}\right)
+e_{4}\left(\left(\sqrt{g^{-1}f}\right)^{\mu}~_{\nu}\right).
\end{equation}
Considering non-minimal models leads to quite complicated calculations, whereas in the minimal model the interaction term of two metrics $g_{\mu\nu}$ and $f_{\mu\nu}$ is just obtained by the trace of $\left(\sqrt{g^{-1}f}\right)^{\mu}~_{\nu}$. It should be mentioned that this simplification does not absolutely change our following results. Before starting the next section, it would be worthwhile to explain a little about the details of extracting the field equations of bigravity model. The variation of action (\ref{Fbi1}) is given by
\be
\label{Fbi5}
\delta_{g}S_{{\rm bi}}=M_{g}^{2}\int d^{4}x\delta_{g} \left(\sqrt{- {\rm det} g}R^{(g)}\right)-2m^{2}M_{{\rm eff}}^{2}\int d^{4}x\delta_{g}\left(\sqrt{- {\rm det} g}\left(3-{\rm tr}\sqrt{g^{-1}f}+{\rm det }\sqrt{g^{-1}f}\right)\right),
\ee
and
\be
\label{Fbi6}
\delta_{f}S_{{\rm bi}}=M_{f}^{2}\int d^{4}x\delta_{f} \left(\sqrt{- {\rm det} f}R^{(f)}\right)-2m^{2}M_{{\rm eff}}^{2}\int d^{4}x\delta_{f}\left(\sqrt{- {\rm det} g}\left(3-{\rm tr}\sqrt{g^{-1}f}+{\rm det }\sqrt{g^{-1}f}\right)\right).
\ee
Clearly, the first terms of two above equations produce the known Einstein tensors for metrics $g_{\mu\nu}$ and $f_{\mu\nu}$, respectively as
\begin{align}
\label{Fbi7}
\delta_{g}\left(\sqrt{- {\rm det} g}R^{(g)}\right)=\sqrt{- {\rm det} g}\left(R_{\mu\nu}^{(g)}-\frac{1}{2}g_{\mu\nu}R^{(g)}\right),
\end{align}
and
\be
\label{Fbi8}
\delta_{f}\left(\sqrt{- {\rm det} f}R^{(f)}\right)=\sqrt{- {\rm det} f}\left(R_{\mu\nu}^{(f)}-\frac{1}{2}f_{\mu\nu}R^{(f)}\right).
\ee
By considering that $\delta {\rm tr}\left(\sqrt{g^{-1}f}\right)=\frac{1}{2} {\rm tr}\left(g\sqrt{g^{-1}f}\delta g^{-1}\right)=\frac{1}{2}{\rm tr}\left(f\left(\sqrt{g^{-1}f}\right)^{-1}\delta g^{-1}\right)$ beside the property $\sqrt{{\rm det}(-g)}\sqrt{{\rm det}g^{-1}f}=\sqrt{{\rm det}(-f)}$ we can conclude that

\begin{align}
\label{Fbi9}
0=&M_{g}^{2}\left(R_{\mu\nu}^{(g)}-\frac{1}{2}R^{(g)}g_{\mu\nu}\right)\nn&+m^{2}M_{{\rm eff}}^{2}\left\{g_{\mu\nu}\left(3-{\rm tr}\sqrt{g^{-1}f}\right)+\frac{1}{2}f_{\mu\rho}\left(\sqrt{g^{-1}f}\right)^{-1 \rho}~_{\nu}+\frac{1}{2}f_{\nu\rho}\left(\sqrt{g^{-1}f}\right)^{-1 \rho}~_{\mu}\right\},
\end{align}
and
\begin{align}\label{Fbi10}
0=&M_{f}^{2}\left(R_{\mu\nu}^{(f)}-\frac{1}{2}R^{(f)}f_{\mu\nu}\right)\nn&+m^{2}M_{{\rm eff}}^{2}\sqrt{{\rm det}\left(f^{-1}g\right)}\left\{-\frac{1}{2}f_{\mu\rho}\left(\sqrt{g^{-1}f}\right)^{ \rho}~_{\nu} -\frac{1}{2}f_{\nu\rho}\left(\sqrt{g^{-1}f}\right)^{ \rho}~_{\mu}+f_{\mu\nu}{\rm det}\left(\sqrt{g^{-1}f}\right)\right\}.
\end{align}
Here, we are ended up \textbf{with} two independent equations of motion for metrics $g_{\mu\nu}$ and $f_{\mu\nu}$ in general model of bigravity.

\section{Bigravity With two Scalar Field Kinetic Terms Non-minimally Coupled to Curvatures \label{Sec3}}

Let us review the original scalar tensor bigravity action

\begin{align}\label{Fbi11}
S_{{\rm tot}}=&M_{g}^{2}\int d^{4}x\sqrt{- {\rm det} g}R^{(g)}+M_{f}^{2}\int d^{4}x\sqrt{- {\rm det} f}R^{(f)}+S_{\phi}+S_{\xi}+\nn&
2m^{2}M_{{\rm eff}}^{2}\int d^{4}x\sqrt{- {\rm det} g}\left(3-{\rm tr}\sqrt{g^{-1}f}+{\rm det }\sqrt{g^{-1}f}\right),
\end{align}
where we have
\be
\label{Fbi12}
S_{\phi}=-M_{g}^{2}\int d^{4}x \sqrt{- {\rm det} g} \left\{\epsilon g_{\mu\nu} \partial^{\mu}\phi \partial^{\nu}\phi+2V(\phi)\right\}+\int d^{4}x {\cal{L}}_{{\rm matter}}\\(g_{\mu\nu},\Phi_{i}),
\ee
and
\be
\label{Fbi13}
S_{\xi}=-M_{f}^{2}\int d^{4}x \sqrt{- {\rm det} f} \left\{\lambda f_{\mu\nu} \partial^{\mu}\xi \partial^{\nu}\xi+2U(\xi)\right\},
\ee
where $V(\phi)$ and $U(\xi)$ are scalar field potentials of metrics $g_{\mu\nu}$ and $f_{\mu\nu}$, respectively. Now, we are interested in writing the modified form of the above action by means of including the non-minimal kinetic derivative couplings to the curvatures in the action as follows

\begin{align}\label{Fbi14}
S_{{\rm tot}}=&M_{g}^{2}\int d^{4}x\sqrt{- {\rm det} g}\left(R^{(g)}-\left[\varepsilon g_{\mu\nu}+\kappa G^{(g)}_{\mu\nu}\right]\partial^{\mu}\phi \partial^{\nu}\phi-2V(\phi)\right)+\nn&M_{f}^{2}\int d^{4}x\sqrt{- {\rm det} f}\left(R^{(f)}-\left[\lambda f_{\mu\nu}+\beta G^{(f)}_{\mu\nu}\right]\partial^{\mu}\xi \partial^{\nu}\xi-2U(\xi)\right)+\nn&2m^{2}M_{{\rm eff}}^{2}\int d^{4}x\sqrt{- {\rm det} g}\left(3-{\rm tr}\sqrt{g^{-1}f}+{\rm det }\sqrt{g^{-1}f}\right),
\end{align}
where we have ignored the matter contribution, and $\kappa$ and $\lambda$ are the $g_{\mu\nu}$ and  $f_{\mu\nu}$ coupling parameters with dimension of $(\textit{length})^{2}$, respectively. Varying the action (\ref{Fbi14}) with respect to $g_{\mu\nu}$ and $\phi$ yields the field equations

\begin{align}\label{Fbi15}
0=&-G^{(g)}_{\mu\nu}+M_{g}^{2}T^{(\phi)}_{\mu\nu}+\kappa M_{g}^{2}\Theta^{(g)}_{\mu\nu}+\nn&m^{2}M_{{\rm eff}}^{2}\left\{g_{\mu\nu}\left(3-{\rm tr}\sqrt{g^{-1}f}\right)
+\frac{1}{2}f_{\mu\rho}\left(\sqrt{g^{-1}f}\right)^{-1 \rho}~_{\nu}
+\frac{1}{2}f_{\nu\rho}\left(\sqrt{g^{-1}f}\right)^{-1 \rho}~_{\mu}\right\},
\end{align}
and
\be\label{Fbi16}
0=\left[\varepsilon g^{\mu\nu}+\kappa G^{(g)}_{\mu\nu}\right]\nabla_{\mu}\nabla_{\nu}\phi -V_{\phi},
\ee
respectively, where $V_{\phi}\equiv dV(\phi)/ d\phi$, $\nabla^{\mu}\equiv\nabla^{\mu}_{g}$, and
\be\label{Fbi17}
T^{(\phi)}_{\mu\nu}=\varepsilon\left[\nabla_{\mu}\phi \nabla_{\nu}\phi-\frac{1}{2}g_{\mu\nu}(\nabla \phi)^{2}\right]-g_{\mu\nu}V(\phi),
\ee

\begin{align}\label{Fbi18}
\Theta^{(g)}_{\mu\nu}=&-\frac{1}{2}\nabla_{\mu}\phi \nabla_{\nu}\phi R^{(g)}+2\nabla_{\alpha}\phi \nabla_{(\mu}\phi R^{(g)\alpha}~_{\nu )}+\nabla^{\alpha}\phi \nabla^{\beta}\phi R^{(g)}_{\mu\alpha\nu\beta}+\nn&
\nabla_{\mu}\nabla^{\alpha}\phi \nabla_{\nu}\nabla_{\alpha}\phi-\nabla_{\mu}\nabla_{\nu}\phi \Box\phi-\frac{1}{2}G^{(g)}_{\mu\nu}(\nabla \phi)^{2}
+\nn&
g_{\mu\nu}\left[-\frac{1}{2}\nabla^{\alpha} \nabla^{\beta}\phi\nabla_{\alpha}\phi \nabla_{\beta}\phi+\frac{1}{2}\left(\Box\phi\right)^{2}
-\nabla_{\alpha}\phi \nabla_{\beta}R^{(g)\alpha\beta}\right],
\end{align}
where $\Box\equiv\Box_{g}$ is the d'Alembertian with respect to the metric $g$. Using (\ref{Fbi14}), and because $f$ is dynamical as well as $g$, there is a symmetry between them so that we are allowed to write
\begin{align}\label{Fbi19}
0=&-G^{(f)}_{\mu\nu}+M_{f}^{2}T^{(\xi)}_{\mu\nu}+\beta M_{g}^{2}\Theta^{(f)}_{\mu\nu}+\nn&
m^{2}M_{{\rm eff}}^{2}\sqrt{{\rm det}\left(f^{-1}g\right)}\left\{-\frac{1}{2}f_{\mu\rho}\left(\sqrt{g^{-1}f}\right)^{ \rho}~_{\nu} -\frac{1}{2}f_{\nu\rho}\left(\sqrt{g^{-1}f}\right)^{ \rho}~_{\mu}+f_{\mu\nu}{\rm det}\left(\sqrt{g^{-1}f}\right)\right\},
\end{align}

\be\label{Fbi20}
0=\left[\lambda f^{\mu\nu}+\beta G^{(f)}_{\mu\nu}\right]\nabla_{\mu}\nabla_{\nu}\phi -U_{\xi},
\ee
where $U_{\xi}\equiv dU(\xi)/d\xi$. Under the mentioned symmetry condition $\nabla^{\mu}_{g}\leftrightarrow \nabla^{\mu}_{f}$ and $\Box_{g}\leftrightarrow \Box_{f}$ we just have the  changes
\be\label{Fbi21}
T^{(\xi)}_{\mu\nu}=\lambda\left[\nabla_{\mu}\xi \nabla_{\nu}\xi-\frac{1}{2}f_{\mu\nu}(\nabla \xi)^{2}\right]-f_{\mu\nu}U(\xi),
\ee

\begin{align}\label{Fbi22}
\Theta^{(f)}_{\mu\nu}=&-\frac{1}{2}\nabla_{\mu}\xi \nabla_{\nu}\xi R^{(f)}+2\nabla_{\alpha}\xi \nabla_{(\mu}\xi R^{(f)\alpha}~_{\nu )}+\nabla^{\alpha}\xi \nabla^{\beta}\xi R^{(f)}_{\mu\alpha\nu\beta}+\nn&
\nabla_{\mu}\nabla^{\alpha}\xi \nabla_{\nu}\nabla_{\alpha}\xi-\nabla_{\mu}\nabla_{\nu}\xi \Box\xi-\frac{1}{2}G^{(f)}_{\mu\nu}(\nabla \xi)^{2}
+
f_{\mu\nu}\left[-\frac{1}{2}\nabla^{\alpha} \nabla^{\beta}\xi\nabla_{\alpha}\xi \nabla_{\beta}\xi+\frac{1}{2}\left(\Box\xi\right)^{2}
-\nabla_{\alpha}\xi\nabla_{\beta}R^{(f)\alpha\beta}\right].
\end{align}

According to Bianchi identity we have

\be\label{Fbi23}
0=\nabla^{\mu}_{g}\left(R_{\mu\nu}^{(g)}-\frac{1}{2}g_{\mu\nu}R^{(g)}\right),
\ee

\be\label{Fbi24}
0=\nabla^{\mu}_{f}\left(R_{\mu\nu}^{(f)}-\frac{1}{2}f_{\mu\nu}R^{(f)}\right).
\ee

By imposing the covariant derivative $\nabla^{\mu}_{g}$ and $\nabla^{\mu}_{f}$ on equations (\ref{Fbi15}) and (\ref{Fbi19}) and using the constraints $\nabla^{\mu}\left(M_{g}^{2}T^{(\phi)}_{\mu\nu}+\kappa M_{g}^{2}\Theta^{(g)}_{\mu\nu}\right)=0$, and $\nabla^{\mu}\left(M_{f}^{2}T^{(\xi)}_{\mu\nu}+\beta M_{f}^{2}\Theta^{(f)}_{\mu\nu}\right)=0,$ resulting from equations (\ref{Fbi16}), (\ref{Fbi17}), (\ref{Fbi18}), (\ref{Fbi20}), (\ref{Fbi21}) and (\ref{Fbi22}), besides refereing to (\ref{Fbi23}) and (\ref{Fbi24}), we can obtain two following constraints

\be\label{Fbi25}
0=-g_{\mu\nu}\nabla^{\mu}_{g}\left(tr\sqrt{g^{-1}f}\right)+\frac{1}{2}\nabla^{\mu}_{g}\left\{f_{\mu\rho}\left(\sqrt{g^{-1}f}\right)^{-1 \rho}~_{\nu}+f_{\nu\rho}\left(\sqrt{g^{-1}f}\right)^{-1 \rho}~_{\mu}\right\},
\ee

\be\label{Fbi26}
0=\nabla^{\mu}_{f}\left[\sqrt{{\rm det}(f^{-1}g)}\left\{-\frac{1}{2}\left(\sqrt{g^{-1}f}\right)^{-1\nu}~_{\sigma}g^{\sigma\mu}-
\frac{1}{2}\left(\sqrt{g^{-1}f}\right)^{-1\mu}~_{\sigma}g^{\sigma\nu}+f^{\mu\nu}{\rm det}\left(\sqrt{g^{-1}f}\right)\right\}\right],
\ee
which can be used to extract an important constraint on the metric coefficients implying that we can find a dynamical cosmology in this model.

\section{Cosmological Equations with  Non-minimal Kinetic Coupling bigravity model\label{Sec4}}

As an important step to test a theory as a real cosmological model, it would be fruitful to inquire whether it is possible to have a model expressing the arbitrary evolution of the expanding universe. As a result, now we consider the FRW universe for the $g_{\mu\nu}$ metric besides using the conformal time $t=\tau$. We  emphasize that in bigravity not only the metric  $g_{\mu\nu}$ but also $f_{\mu\nu}$ is dynamical, therefore we take the following form of metrics

\begin{equation}\label{Fbi27}
ds_{g}^{2}=\sum_{\mu,\nu=0}^{3}g_{\mu\nu}dx^{\mu}dx^{\nu}=a(\tau)^{2}\left(-d\tau^{2}+\sum_{i=1}^{3}(dx^{i})^{2}\right),
\end{equation}
\begin{equation}\label{Fbi28}
ds_{f}^{2}=\sum_{\mu,\nu=0}^{3}f_{\mu\nu}dx^{\mu}dx^{\nu}=-c(\tau)^{2}d\tau^{2}+b(\tau)^{2}\sum_{i=1}^{3}(dx^{i})^{2}.
\end{equation}

Obviously we are not allowed to take $c(\tau)=1$ nor $c(\tau)=b(\tau)$, because by this option we will have the Minkowski metric for $ds_{f}^{2}$ that leads to Massive gravity non-dynamical cosmology \cite{30}. Thus, for this case, the $(\tau,\tau)$ and $(i,i)$ components of (\ref{Fbi15}) are given by

\be\label{Fbi29}
0=-3M_{g}^{2}H^{2}-3m^{2}M_{{\rm eff}}^{2}(a^{2}-ab)+\left(\frac{\varepsilon\dot{\phi}^{2}}{2}+
V(\phi)a(\tau)^{2}\right)M_{g}^{2}-9\kappa M_{g}^{2}\frac{ H^{2}\dot{\phi}^{2}}{2a^{2}},
\ee

\begin{align}\label{Fbi30}
0=&M_{g}^{2}\left(2\dot{H}+H^{2}\right)+m^{2}M_{{\rm eff}}^{2}(3a^{2}-2ab-ac)+\left(\frac{\varepsilon\dot{\phi}^{2}}{2}-
V(\phi)a(\tau)^{2}\right)M_{g}^{2}-\nn&
\kappa M_{g}^{2}\left[\frac{\dot{H}\dot{\phi}^{2}}{a^{2}}-\frac{3H^{2}\dot{\phi}^{2}}{2a^{2}}+\frac{2H\dot{\phi}\ddot{\phi}}{a^{2}}\right].
\end{align}
 Moreover, the $(\tau,\tau)$ and $(i,i)$ components of (\ref{Fbi19}) yield

\be\label{Fbi301}
0=-3M_{f}^{2}K^{2}+m^{2}M_{{\rm eff}}^{2}c^{2}(1-\frac{a^{3}}{b^{3}})+\left(\frac{\lambda \dot{\xi}^{2}}{2}+
U(\xi)c(\tau)^{2}\right)M_{f}^{2}-9\beta M_{f}^{2}\frac{ K^{2}\dot{\xi}^{2}}{2a^{2}},
\ee

\begin{align}\label{Fbi31}
0=&M_{f}^{2}\left(-2\dot{K}-3K^{2}+2KL\right)+m^{2}M_{{\rm eff}}^{2}(-\frac{a^{3}c}{b^{2}}+c^{2})+\left(-\frac{\lambda\dot{\xi}^{2}}{2}+
U(\xi)a(\tau)^{2}\right)M_{f}^{2}+\nn&
\frac{\beta M_{f}^{2}}{c^{2}}\left[\dot{\xi}^{2}\left(-\dot{K}-\frac{3K^{2}}{2}+3KL\right)-2K\dot{\xi}\ddot{\xi}\right],
\end{align}
by definition of $K=\dot{b}/b$ and $L=\dot{c}/c$. Applying two scalar fields for metrics  (\ref{Fbi27}) and (\ref{Fbi28}) helps us to describe three metric coefficients $a(\tau)$, $b(\tau)$ and $c(\tau)$ as three degrees of freedom;  this is not possible just by one scalar field.
It turns out  that the equations   (\ref{Fbi25})
and (\ref{Fbi26}) carry  important results for the variables defined in the
metrics (\ref{Fbi27})
and (\ref{Fbi28}) as follows
\be\label{Fbi32}
cH=bK~~{\rm or}~\frac{c\dot{a}}{a}=\dot{b}.
\ee

This is a constraint imposing on the metrics and relating them to each other. For the case $\dot{a}\neq0$ the above constraint gives $c=a\dot{b}/\dot{a}$ but for $\dot{a}=0$, we have $\dot{b}=0$ meaning that $a$ and $b$ are constant but $c$ is arbitrary. Actually, this constraint is allowing us to go on and construct an expanding cosmology. Redefining scalar fields in accordance with conformal times $\eta=\zeta=\tau$ as $\varphi=\varphi(\eta)$ and $\xi=\xi(\zeta)$ with $\omega(\eta)=\varphi'(\eta)^{2}$, $\tilde{V}(\eta)=V\left(\varphi(\eta)\right)$, $\sigma(\zeta)=\xi'(\zeta)^{2}$ (prime is the derivative with respect to its conformal time) and $\tilde{U}(\zeta)=U\left(\xi(\zeta)\right)$ facilitate us to rewrite equations (\ref{Fbi29})-(\ref{Fbi31}) as follows

\be\label{Fbi33}
0=-3M_{g}^{2}H^{2}-3m^{2}M_{{\rm eff}}^{2}(a^{2}-ab)+\left(\frac{\varepsilon \omega(\tau)}{2}+
V(\tau)a(\tau)^{2}\right)M_{g}^{2}-9\kappa M_{g}^{2}\frac{ H^{2}\omega(\tau)}{2a^{2}},
\ee

\begin{align}\label{Fbi34}
0=&M_{g}^{2}\left(2\dot{H}+H^{2}\right)+m^{2}M_{{\rm eff}}^{2}(3a^{2}-2ab-ac)+\left(\frac{\epsilon \omega(\tau)}{2}-
V(\tau)a(\tau)^{2}\right)M_{g}^{2}-\nn&
\frac{\kappa M_{g}^{2}}{a^{2}}\left[\left(\dot{H}-3H^{2}\right)\omega(\tau)+H\dot{\omega}(\tau)\right],
\end{align}

\be\label{Fbi35}
0=-3M_{f}^{2}K^{2}+m^{2}M_{{\rm eff}}^{2}c^{2}(1-\frac{a^{3}}{b^{3}})+\left(\frac{\lambda \sigma(\tau)}{2}+
U(\tau)c(\tau)^{2}\right)M_{f}^{2}-9\beta M_{f}^{2}\frac{ K^{2}\sigma(\tau)}{2a^{2}},
\ee

\begin{align}\label{Fbi36}
0=&M_{f}^{2}\left(-2\dot{K}-3K^{2}+2KL\right)+m^{2}M_{{\rm eff}}^{2}(-\frac{a^{3}c}{b^{2}}+c^{2})+\left(-\frac{\lambda\sigma(\tau)}{2}+
U(\tau)a(\tau)^{2}\right)M_{f}^{2}+\nn&
\frac{\beta M_{f}^{2}}{c^{2}}\left[\sigma(\tau)\left(-\dot{K}-\frac{3K^{2}}{2}+3KL\right)-K\dot{\sigma}(\tau)\right].
\end{align}

Adding and subtracting equations (\ref{Fbi33}) and (\ref{Fbi34}) give us two following equations

\be\label{Fbi37}
0=2M_{g}^{2}\left(\dot{H}-H^{2}\right)+m^{2}M_{{\rm eff}}^{2}(ab-ac)+\epsilon \omega(\tau)
M_{g}^{2}+\frac{\kappa M_{g}^{2}}{a^{2}}\left[\omega(\tau)\left(-6H^{2}+\dot{H}\right)+H\dot{\omega}(\tau)\right],
\ee

\be\label{Fbi38}
0=M_{g}^{2}\left(2\dot{H}+4H^{2}\right)+m^{2}M_{{\rm eff}}^{2}(6a^{2}-5ab-ac)-2M_{g}^{2}a(\tau)^{2}\tilde{V}(\tau)+
\frac{\kappa M_{g}^{2}}{a^{2}}\left[\left(\dot{H}+3H^{2}\right)\omega(\tau)+H\dot{\omega}(\tau)\right].
\ee

Again, by  subtracting and adding equations (\ref{Fbi35}) and (\ref{Fbi36}) we will have
\be\label{Fbi39}
0=2M_{f}^{2}\left(-\dot{K}+KL\right)+m^{2}M_{{\rm eff}}^{2}(-\frac{a^{3}c}{b^{2}}+\frac{a^{3}c^{2}}{b^{3}})-\lambda\sigma(\tau)M_{f}^{2}+
\frac{\beta M_{f}^{2}}{c^{2}}\left[\sigma(\tau)\left(-\dot{K}-6K^{2}+3KL\right)-K\dot{\sigma}(\tau)\right],
\ee
\begin{align}\label{Fbi40}
0=&M_{f}^{2}\left(-6K^{2}+2KL-2\dot{K}\right)+m^{2}M_{{\rm eff}}^{2}c^{2}\left(2-\frac{a^{3}}{b^{3}}-\frac{a^{3}}{b^{2}c}\right)+2M_{f}^{2}c(\tau)^{2}\tilde{U}(\tau)+\nn&
\frac{\beta M_{f}^{2}}{c^{2}}\left[\sigma(\tau)\left(-6K^{2}+3KL-\dot{K}\right)-K\dot{\sigma}(\tau)\right].
\end{align}
Therefore, for arbitrary metric coefficients $a(\tau)$, $b(\tau)$ and $c(\tau)$,
 by choosing $\omega(\tau)$, $\sigma(\tau)$, $\tilde{V}(\tau)$ and $\tilde{U}(\tau)$ satisfying equations (\ref{Fbi37})-(\ref{Fbi40}), we can reconstruct cosmological
models with given evolutions of  $a(\tau)$, $b(\tau)$ and $c(\tau)$.\\

\subsection{Conformal description of the power expanding universe}

In the last section, in the equations (\ref{Fbi27}) and (\ref{Fbi28}), we have applied the conformal metric with a conformal time $\tau$ to express the evolution of the scalar field $\phi(\tau)$. In the following, we are going to show how the conformal time can describe the known cosmologies. As a result, again we remark the conformally flat FLRW metric as

\be\label{Fbi41}
ds^{2}=a(\tau)^{2}\left(-d\tau^{2}+\sum_{i=1}^{3}(dx^{i})^{2}\right).
\ee

According to the power expanding universe the scale factor is treating as $a(\tau)=\frac{a_{0}^{n}}{\tau^{n}}$ with $n\neq1$. Considering $n=1$ can explain the  de Sitter universe in the arbitrary model representing dark energy or inflation. Looking back to the power expanding scale factor leads to the following redefinition of the time coordinate as

\be\label{Fbi42}
d\tilde{t}=\pm \frac{a_{0}^{n}}{\tau^{n}} d\tau,
\ee

so we have

\be\label{Fbi43}
\tilde{t}=\pm \frac{a_{0}^{n}}{n-1}\tau^{1-n}.
\ee

By considering the above discussion, we will have the following form for the FLRW metric

\be\label{Fbi44}
ds^{2}=-d\tilde{t}^{2}+\left[\pm (n-1)\frac{\tilde{t}}{a_{0}}\right]^{\frac{-2n}{1-n}}\sum_{i=1}^{3}(dx^{i})^{2}.
\ee
From this equation one finds that for the range $0<n<1$, the metric describes the Phantom universe \cite{31}, for $n>1$  it describes the quintessence universe and for $n<0$ it describes the decelerating universe. For the case of Phantom universe, we  choose $+$ sign of (\ref{Fbi42}) or (\ref{Fbi43}) and impose the time shift $\tilde{t}\rightarrow \tilde{t}-t_{0}$ \textbf{on} (\ref{Fbi44}).
 By this shift, we have the Big Rip at  $\tilde{t}=t_{0}$, the present time at  $\tilde{t}<t_{0}$, and  the infinite past ($\tilde{t}\rightarrow-\infty$) is equivalent to the limit of $\tau\rightarrow\infty$. {For the quintessence range ($n>1$) we  again choose $+$ sign of (\ref{Fbi42}) or (\ref{Fbi43}).  Hence, the limit of $\tau\rightarrow 0$ corresponds to  $\tilde{t}\rightarrow +\infty$ and the limit of $\tau\rightarrow +\infty$ corresponds to  $\tilde{t}\rightarrow 0$, which may describe the Big Bang}. {Similarly, for the decelerating universe we choose $+$ sign. As a result, the limit of $\tau\rightarrow +\infty$  corresponds to  $\tilde{t}\rightarrow +\infty$, and that of $\tau\rightarrow 0$   corresponds  to $\tilde{t}\rightarrow 0$  which again describes the Big Bang}. Finally, it should be noted that for the de Sitter case $(n=1)$ we have

\begin{align}\label{Fbi45}
if ~\tau\rightarrow 0~~~~~\tilde{t}\rightarrow +\infty~~~~~~~~~~~~~~~~~~~~~~{\rm and}~~&&if ~\tau\rightarrow \pm\infty~~~~~\tilde{t}\rightarrow -\infty.
\end{align}

\subsection{Dark energy solution with $a(\tau)=b(\tau)=c(\tau)$}

 Our universe can be described by the metric $g_{\mu\nu}$. As a result, we have freedom to choose the functions $c(\tau)$ and $b(\tau)$ which are not directly describing the expansion of the universe since they are $f_{\mu\nu}$ degrees of freedom in the Einstein frame.
Any way, we choose the functions  $c(\tau)$ and $b(\tau)$ equal to $a(\tau)$ to have more convenient calculation, however, we should not deduce that they do not have any physical meaning. Thus, we are allowed to take $a(\tau)=b(\tau)=c(\tau)$ leading to $K=H=L$. As a result, the metric interaction terms in equations (\ref{Fbi37})-(\ref{Fbi40}) vanish and also by supposing $M_{f}=M_{g}$, $\epsilon=\lambda $ and $\kappa=\beta$ we can obtain

\be\label{Fbi46}
\kappa \dot{\omega}(\tau)+\omega(\tau)\left(\frac{a(\tau)^{2}\epsilon}{H}+\kappa \left(-6H+\frac{\dot{H}}{H}\right)\right)+2a(\tau)^{2}\left(\frac{\dot{H}}{H}-H\right)=0,
\ee

\be\label{Fbi47}
a(\tau)^{2}\tilde{V}(\tau)=3H^{2}+\omega(\tau)\left(\frac{9\kappa H^{2}}{2a(\tau)^{2}}-\frac{\epsilon}{2}\right),
\ee

\be\label{Fbi48}
\kappa\dot{\sigma}(\tau)+\sigma(\tau)\left(\frac{a(\tau)^{2}\epsilon}{H}+\kappa \left(-6H+\frac{\dot{H}}{H}\right)\right)+2a(\tau)^{2}\left(\frac{\dot{H}}{H}-H\right)=0,
\ee
and
\be\label{Fbi49}
a(\tau)^{2}\tilde{U}(\tau)=3H^{2}+\sigma(\tau)\left(\frac{9\kappa H^{2}}{2a(\tau)^{2}}-\frac{\epsilon}{2}\right).
\ee

Obviously, we can deduce that $\sigma(\tau)=\omega(\tau)$ and so $\tilde{V}(\tau)=\tilde{U}(\tau)$.
By inserting $a(\tau)=\frac{a_{0}^{n}}{\tau^{n}}$ and $a(\tau)=e^{n\tau}$ into (\ref{Fbi46}) and (\ref{Fbi47}) and supposing that $a_{0}=1$, $\epsilon=\frac{1}{2}$, $\kappa=1$ and $\sigma(\tau)=\omega(\tau)$ we can find

\be\label{Fbi50}
\dot{\omega}(\tau)+\omega(\tau)\left(\frac{-1}{2n\tau^{2n-1}}+\frac{6n-1}{\tau}\right)+2\tau^{2n-1}\left(n-1\right)=0,
\ee

 \be\label{Fbi51}
\tilde{V}(\tau)=3n^{2}\tau^{2n-2}+\omega(\tau)\left(\frac{9}{2}n^{2}\tau^{4n-2}-\frac{\tau^{2n}}{4}\right),
\ee

  \be\label{Fbi52}
\dot{\omega}(\tau)+\omega(\tau)\left(\frac{e^{2n\tau}}{n}-6n\right)-2ne^{2n\tau}=0,
\ee

  \be\label{Fbi53}
\tilde{V}(\tau)=3n^{2}e^{-2n\tau}+\omega(\tau)\left(\frac{9}{2}n^{2}e^{-2n\tau}-1\right).
\ee

Clearly, $\omega(\tau)$ and $\tilde{V}(\tau)$ are indirectly coupled to each other, so the $\omega$'s which are obtained from equations (\ref{Fbi50}) and (\ref{Fbi52}) are influenced by the evolution of potential $\tilde{V}(\tau)$. In order to extract the cosmological constant in our model, we need that $\tilde{V}(\tau)$ become constant. Then, by differentiation  both sides of (\ref{Fbi51}) with respect to $\tau$ we can extract the expression for $\dot{\omega}(\tau)$ satisfying that of equation (\ref{Fbi50}). Thus, we are facilitated to write $\omega(\tau)$ for the case $a(\tau)=\frac{1}{\tau^{n}}$ and $\tilde{V}(\tau)={\rm cte}$ as

\begin{align}\label{Fbi54}
\omega(\tau)=\frac{\left(2(1-n)\tau^{2n-1}\right)\left(\frac{9}{2}n^{2}\tau^{2n+1}-
\frac{\tau^{3}}{4}\right)+6n^{2}(n-1)}{\left(\frac{9}{2}n^{2}\tau^{2n+1}-\frac{\tau^{3}}{4}\right)
\left(\frac{9n^{2}\left(1-2n\right)\tau^{4n-3}+\frac{n\tau^{2n-1}}{2}}{\frac{9}{2}n^{2}\tau^{4n-2}-
\frac{\tau^{2n}}{4}}-\frac{1}{2n\tau^{2n-1}}+\frac{6n-1}{\tau}\right)}.
\end{align}

Considering the above result besides referring to (\ref{Fbi51}) makes it clear that $\omega(\tau)$ vanishes and also $\tilde{V}(\tau)$ becomes constant, provided that $n=1$. This result is exactly the same as that of scalar tensor bigravity model explaining the de Sitter model with $a(\tau)=\frac{1}{\tau}$ as a dark energy universe. In the next section, we will seek the stability of all mentioned solutions, even the de Sitter one, under the homogeneous perturbation.

Now, we consider the scale factor $a(\tau)=e^{n\tau}$. Similar to the previous approach for constant potentials we have

 \be\label{Fbi55}
\omega(\tau)=\frac{6n^{3}e^{-1}+2ne^{2n\tau}\left(1-\frac{9}{2}n^{2}e^{-2n\tau}\right)}{\left(6n-
\frac{e^{2n\tau}}{2n}\right)\left(\frac{9}{2}n^{2}e^{-2n\tau}-1\right)+2ne^{-1}-18n^{3}e^{-1-2n\tau}},
\ee
which should take just positive values not to conflict  with the definition $\omega(\eta)=\varphi'(\eta)^{2}$.
Studying the three dimensional diagram of the above equation for $\omega(\tau)$ with the variables $\tau$ and $n$ lets us to write

  \be\label{Fbi56}
\omega(\tau)>0~~~~~\rightarrow~~~~~~\left(100<\tau<+\infty~~~~~~~~~{\rm for~ all ~values~ of~ n}\right).
\ee
Thus, to be more precise, we may describe the cosmological constant solution for the case $a(\tau)=e^{n\tau}$ as follows
\begin{align}\label{Fbi57}
&{\rm for}~~\left(n>0~,\tau\rightarrow +\infty\right)~~{\rm we~have}~~\left(\tilde{V_{0}}\rightarrow 18n^{3}~{\rm and}~\omega(\tau)\rightarrow 4n\right),\nn& {\rm for}~~\left(n<0~,\tau\rightarrow +\infty\right)~~{\rm we~have}~~\left(\tilde{V_{0}}\rightarrow0~{\rm and}~\omega(\tau)\rightarrow 0\right).
\end{align}

\section{Cosmology and Stability of Solutions\label{Sec5}}

Following the last section, we shall discuss about the validity of these wide range of presented solutions to choose just the cases which satisfy both the consistency condition for our model and the stability constraint.  In the following, we want to investigate the stability of the solutions discussed in the previous section.

 Going through equations (\ref{Fbi46}) and (\ref{Fbi48}), shows us a way to find the positive ranges of $\omega(\tau)$ and $\sigma(\tau)$ by means of plotting their three dimensional diagrams  extracted by equations (\ref{Fbi46}) and (\ref{Fbi48}) in which we put two mentioned scale factors $a(\tau)=\frac{a_{0}^{n}}{\tau^{n}}$ and $a(\tau)=e^{n\tau}$ . As we know,  $\omega(\tau)$ and $\sigma(\tau)$ are $\dot{\phi}^{2}$ and $\dot{\xi}^{2}$, respectively, so they are not allowed to take negative values in order to avoid of  ghost and inconsistency in the theory.\\
By inserting $a(\tau)=\frac{a_{0}^{n}}{\tau^{n}}$ and $a(\tau)=e^{n\tau}$ into (\ref{Fbi46}) and supposing that $a_{0}=1$, $\epsilon=\frac{1}{2}$, $\kappa=1$ and $\sigma(\tau)=\omega(\tau)$ we can find

\be\label{Fbi58}
\dot{\omega}(\tau)+\omega(\tau)\left(\frac{-1}{2n\tau^{2n-1}}+\frac{6n-1}{\tau}\right)+2\tau^{2n-1}\left(n-1\right)=0,
\ee

\be\label{Fbi59}
\dot{\omega}(\tau)+\omega(\tau)\left(\frac{e^{2n\tau}}{n}-6n\right)-2ne^{2n\tau}=0,
\ee
respectively. Two above equations have the following solutions
\begin{align}\label{Fbi60}
\omega(\tau)=&e^{\frac{1}{2} \left(\frac{\tau^{2-2 n}}{(2-2 n) n}-2 (6 n-1) \ln (\tau)\right)} c_1-2 e^{\frac{1}{2} \left(\frac{\tau^{2-2 n}}{(2-2 n) n}-2 (6 n-1) \ln (\tau)\right)} (n-1)\times\nn&
   \left(\frac{2^{\frac{5 n+2}{1-n}} \tau^7 \Gamma \left(\frac{7}{2-2 n},\frac{\tau^{2-2 n}}{4 n-4 n^2}\right) \left(\frac{\tau^{2-2 n}}{n-n^2}\right)^{\frac{7}{2
   (n-1)}}}{n^4 \left(384 n^5+880 n^4-500 n^3-760 n^2-19 n+15\right)}+ \right.\nn
& \left.       \frac{e^{\frac{\tau^{2-2 n}}{4 (n-1) n}} \tau^{2 n-1} \left(8 n^3 \left(48 n^3+164 n^2+116
   n+15\right) \tau^{6 n}+2 n (2 n+5) \tau^{2 n+4}+4 n^2 \left(8 n^2+26 n+15\right) \tau^{4 n+2}+\tau^6\right)}{8 n^3 \left(384 n^4+1264 n^3+764 n^2+4 n-15\right)}\right),
\end{align}

\begin{align}\label{Fbi61}
\omega(\tau)=e^{6 n \tau-\frac{e^{2 n \tau}}{4 n^2}} c_2+2 e^{6 n \tau-\frac{e^{2 n \tau}}{4 n^2}} n \left(e^{\frac{e^{2 n \tau}}{4 n^2}} \left(-\frac{e^{-4 n \tau}}{4 n}-\frac{e^{-2 n \tau}}{16
   n^3}\right)+\frac{\text{Ei}\left(\frac{e^{2 n \tau}}{4 n^2}\right)}{64 n^5}\right),
\end{align}
where $c_{1}$ and $c_{2}$ are the integration constants.
The allowed ranges of $\tau$ and $n$ resulting in the positive valued  $\omega$
are given by the Table.1 and Table.2\footnote{To find the positive valued  $\omega$, we first plotted $\omega(\tau, n)$ in the three dimensional diagrams with respect to $\tau$  (for three ranges of conformal time within two classes  of $\tau>0$ and $\tau<0$) and $n$ (for three ranges  given in subsection A). Then, we  selected the allowed ranges of $\tau$ and $n$ resulting in positive $\omega$.}.

 \vspace{5mm}
\begin{center}
{\scriptsize{ Table 1: }}\hspace{-2mm} {\scriptsize  Intervals for $\tau$ and $n$ resulting from the condition $\omega(\tau)>0$ for the case $a(\tau)=\frac{1}{\tau^{n}}$.}\\
    \begin{tabular}{|l| l |l |  p{800mm} }
    \hline
   {\footnotesize$~~~~~~\tau$ }& ~~{\footnotesize~~~~~ $n$ }  \\ \hline
{\footnotesize ~$\left(0,10^{-20}\right)$} & ~~{\footnotesize $\left(0.38,0.42\right)$} \\\hline
 {\footnotesize ~~~~$\left(0,1\right)$} & ~~{\footnotesize  ~$\left(1,+\infty\right)$}
\\ \hline
{\footnotesize ~~$\left(1,10^{4}\right)$} & ~~{\footnotesize  ~~~$\left(0,1\right)$}\\ \hline
{\footnotesize ~$\left(10^{4},+\infty\right)$} & ~~{\footnotesize  $\left(0.7,0.9\right)$}\\ \hline

    \end{tabular}
\end{center}

 \vspace{5mm}
\begin{center}
{\scriptsize{ Table 2: }}\hspace{-2mm} {\scriptsize  Intervals for $\tau$ and $n$ resulting from the condition $\omega(\tau)>0$ for the case $a(\tau)=e^{n\tau}$.}\\
    \begin{tabular}{|l| l |l |  p{133mm} }
    \hline
   {\footnotesize$~~~~~~\tau$ }& ~~{\footnotesize~~~~~ $n$ }  \\ \hline
{\footnotesize ~~$\left(-1,0\right)$} & ~{\footnotesize $\left(-\infty,0.25\right)$}
\\\hline
 {\footnotesize ~$\left(-12,-10\right)$} & ~{\footnotesize  ~$\left(-0.57,0\right)$}
\\ \hline
{\footnotesize ~~$\left(-0.06,0\right)$} & ~~{\footnotesize  $\left(1,+\infty\right)$}
\\ \hline
{\footnotesize ~~~~$\left(0,1\right)$} & ~{\footnotesize  ~$\left(-0.3,0\right)$}
\\ \hline
{\footnotesize ~~~$\left(1,10\right)$} & ~{\footnotesize  $\left(0.2,0.73\right)$}
\\ \hline
{\footnotesize ~~~~$\left(0,1\right)$} & ~{\footnotesize  $\left(0.1,0.9\right)$}
\\ \hline
{\footnotesize ~~~~$\left(0,1\right)$} & ~{\footnotesize  $\left(1,+\infty\right)$}
\\ \hline
{\footnotesize ~~~~$\left(100,+\infty\right)$} & ~{\footnotesize  $\left(-\infty,+\infty\right)$}
\\ \hline
    \end{tabular}
\end{center}
Referring to equations  (\ref{Fbi33})- (\ref{Fbi36}) and rewriting them by including the conformal times $\eta$ and $\zeta$, gives us

\begin{align}\label{Fbi62}
0=&2M_{g}^{2}\left(\dot{H}-H^{2}\right)+m^{2}M_{{\rm eff}}^{2}\left(a(\tau)b(\tau)-a(\tau)c(\tau)\right)+\epsilon \omega(\eta)\dot{\eta}^{2}
M_{g}^{2}+\nn&\frac{\kappa M_{g}^{2}}{a(\tau)^{2}}\left[\omega(\eta)\dot{\eta}^{2}\left(-6H^{2}+\dot{H}\right)+H\omega'(\eta)\dot{\eta}^{3}
+2H\omega(\eta)\dot{\eta}\ddot{\eta}\right],
\end{align}

\begin{align}\label{Fbi63}
0=&M_{g}^{2}\left(2\dot{H}+4H^{2}\right)+m^{2}M_{{\rm eff}}^{2}\left(6a(\tau)^{2}-5a(\tau)b(\tau)-a(\tau)c(\tau)\right)-2M_{g}^{2}a(\tau)^{2}\tilde{V}(\eta)+\nn&
\frac{\kappa M_{g}^{2}}{a(\tau)^{2}}\left[\left(\dot{H}+3H^{2}\right)\omega(\eta)\dot{\eta}^{2}+H\omega'(\eta)\dot{\eta}^{3}
+2H\omega(\eta)\dot{\eta}\ddot{\eta}\right],
\end{align}

\begin{align}\label{Fbi64}
0=&2M_{f}^{2}\left(-\dot{K}+KL\right)+m^{2}M_{{\rm eff}}^{2}\left(-\frac{a(\tau)^{3}c(\tau)}{b(\tau)^{2}}+\frac{a(\tau)^{3}c(\tau)^{2}}{b(\tau)^{3}}\right)-\lambda\sigma(\zeta)\dot{\zeta}^{2}M_{f}^{2}+
\nn&
\frac{\beta M_{f}^{2}}{c(\tau)^{2}}\left[\sigma(\zeta)\dot{\zeta}^{2}\left(-\dot{K}-6K^{2}+3KL\right)-K\sigma'(\zeta)\dot{\zeta}^{3}-
2K\sigma(\zeta)\dot{\zeta}\ddot{\zeta}\right],
\end{align}

\begin{align}\label{Fbi65}
0=&M_{f}^{2}\left(-6K^{2}+2KL-2\dot{K}\right)+m^{2}M_{{\rm eff}}^{2}c(\tau)^{2}\left(2-\frac{a(\tau)^{3}}{b(\tau)^{3}}-\frac{a(\tau)^{3}}{b(\tau)^{2}c(\tau)}\right)+
2M_{f}^{2}\tilde{U}(\zeta)c(\tau)^{2}+\nn&\frac{\beta M_{f}^{2}}{c(\tau)^{2}}\left[\sigma(\zeta)\dot{\zeta}^{2}\left(-6K^{2}+3KL-\dot{K}\right)-K\sigma'(\zeta)\dot{\zeta}^{3}-
2K\sigma(\zeta)\dot{\zeta}\ddot{\zeta}\right].
\end{align}

Additionally, we can write the scalar field equations as follows

\begin{align}\label{Fbi66}
0=\omega(\eta)\ddot{\eta}+\frac{\omega'(\eta)\dot{\eta}^{2}}{2}+2H\omega(\eta)\dot{\eta}+\tilde{V}'(\eta)a^{2}-
\frac{3K}{a^{2}}\left[\frac{H^{2}\omega'(\eta) \dot{\eta}^{2}}{2}+H^{2}\omega(\eta)\ddot{\eta}+2H\dot{H}\dot{\eta}\omega(\eta)\right],
\end{align}

and

\begin{align}\label{Fbi67}
0=\tilde{\sigma}(\zeta)\ddot{\zeta}+\frac{\sigma'(\zeta)\dot{\zeta}^{2}}{2}+(3K-L)\sigma(\zeta)\dot{\zeta}+\tilde{U}'(\zeta)c^{2}-\frac{\beta}{c^{2}}\left[3K^{2}\left(\frac{\sigma'(\zeta)\dot{\zeta}^{2}}{2}+\sigma(\zeta)\ddot{\zeta}\right)+
\sigma(\zeta)\dot{\zeta}\left(9K^{3}-9K^{2}L+6K\dot{K}\right)\right].
\end{align}
Equations (\ref{Fbi46}) and (\ref{Fbi47}) will give us the metric coefficients provided that we specify these four equations with a function $f(\tau)$ as

\be\label{Fbi68}
\omega'(\eta)+\omega(\eta)\left[\frac{\epsilon e^{2f(\tau)}}{\kappa f'(\eta)}-6f'(\eta)+\frac{f''(\eta)}{f'(\eta)}\right]+\frac{2(\eta)e^{2f(\tau)}}{\kappa}\left(\frac{f''(\eta)}{f'(\eta)}-f'(\eta)\right)=0,
\ee

\be\label{Fbi69}
\tilde{V}(\eta)=3f'^{2}(\eta)e^{-2f(\tau)}+\omega(\eta)\left[\frac{9}{2}\kappa f'^{2}(\eta)e^{-4f(\tau)}-
\frac{\epsilon}{2}e^{-2f(\tau)}\right],
\ee

\be\label{Fbi70}
\sigma'(\zeta)+\frac{2e^{2f(\zeta)}}{\beta}\left(\frac{f''(\zeta)}{f'(\zeta)}-f'(\zeta)\right)+
\sigma(\zeta)\left[\frac{\lambda e^{2f(\zeta)}}{\beta f'(\zeta)}-6f'(\zeta)+\frac{f''(\zeta)}{f'(\zeta)}\right]=0,
\ee

\be\label{Fbi71}
\tilde{U}(\zeta)=3f'^{2}(\zeta)e^{-2f(\zeta)}+\omega(\zeta)\left[\frac{9}{2}\beta f'^{2}(\zeta)e^{-4f(\zeta)}-
\frac{\lambda}{2}e^{-2f(\zeta)}\right].
\ee
As a result, we have
\be\label{Fbi72}
a(\tau)=b(\tau)=c(\tau)=e^{f(\tau)}~,~~\eta=\zeta=\tau.
\ee
We intend to study the stability of solution (\ref{Fbi72}), so we consider the following perturbation to reconstruct the perturbation equations

\be\label{Fbi73}
\zeta\rightarrow\zeta+\delta \zeta,~~~\eta\rightarrow \eta+\delta \eta,~~~b\rightarrow b\left(1+\delta f_{b}\right),~~~a\rightarrow a\left(1+\delta f_{a}\right),~~~K\rightarrow K+\delta K,~~~H\rightarrow H+\delta H.
\ee
Having more simplicity persuade us to consider the following terms:

\be\label{Fbi74}
M_{f}^{2}=M_{g}^{2}=\frac{M_{{\rm eff}}^{2}}{2}=M^{2},
\ee

\begin{align}\label{Fbi75}
\epsilon=\lambda=1/2 ~~~~,~~~~\kappa=\beta.
\end{align}
After a troublesome calculation  explained in Appendix A, we obtain

\begin{align}
\label{Fbibi76}
 \frac{d}{d\tau}\left(
 \begin{array}{c}
 \delta\eta  \\
  \delta H  \\
  \delta\zeta  \\
  \delta K\\
  \delta f_{a}\\
  \delta f_{b}
  \end{array}
  \right)=\left(
  \begin{array}{cccccc}
  -\frac{b_{5}}{b_{4}} & -\frac{b_{3}}{b_{4}} & 0 & 0 & -\frac{b_{1}}{b_{4}}&-\frac{b_{2}}{b_{4}} \\
  -\frac{h_{7}}{h_{1}}+\frac{h_{6}b_{5}}{h_{1}b_{4}} &  -\frac{h_{2}}{h_{1}}+\frac{h_{6}b_{3}}{b_{1}b_{4}} & 0&- \frac{h_{3}}{h_{1}}&-\frac{h_{4}}{h_{1}}+\frac{h_{6}b_{1}}{h_{1}b_{4}}&-\frac{h_{5}}{h_{1}}+\frac{h_{6}b_{2}}{h_{1}b_{4}}\\
  0 & -\frac{c_{4}}{c_{3}}& -\frac{c_{6}}{c_{3}} &-\frac{c_{5}}{c_{3}}&-\frac{c_{1}}{c_{3}}&-\frac{c_{2}}{c_{3}}\\
  g_{5} & g_{1} & g_{6} & g_{2}&g_{4}&g_{3}\\
  0&1&0&0&0&0\\
  0 &0&0&1&0&0\\
  \end{array}
  \right)\left(
  \begin{array}{c}
  \delta\eta  \\
  \delta H  \\
  \delta\zeta  \\
  \delta K\\
  \delta f_{a}\\
  \delta f_{b}
  \end{array}\right),
  \end{align}
where the $(6\times6)$ perturbation matrix  is called ${\rm M}$ and its elements have been defined in Appendix A.  The eigenvalue equation with eigenvalues $\lambda$ has the following form

\be\label{Fbi77}
0=\lambda^{6}+q_{5}\lambda^{5}+q_{4}\lambda^{4}+q_{3}\lambda^{3}+q_{2}\lambda^{2}+q_{1}\lambda+q_{0}.
\ee
Since we are interested in finding stable solutions, we should consider the negative value of the eigenvalues $\lambda$. As a result, all the eigenmodes gradually disappear and thus it ends up in a damped perturbation. Note
that the equation  (\ref{Fbi77}) is respecting the general form of the  eigenvalue function expansion of an arbitrary matrix (${\rm n} \times  {\rm n}$) as
$f_{M}(\lambda)=(-\lambda)^{n}+{\rm tr M}(-\lambda)^{n-1}+...+{\rm detM}$,
which can help us to conclude that this equation with negative eigenvalues requires
\be\label{Fbi78}
q_{5}=-{\rm tr M}>0,
\ee
or (see the Appendix B)
\be\label{Fbi79}
{\rm tr M}=-\frac{h_{2}}{h_{1}}+\frac{h_{6}b_{3}}{b_{1} b_{4}}-\frac{b_{5}}{b_{4}}-\frac{c_{6}}{c_{3}}+g_{2}<0.
\ee
Similar to the previous argument in finding the allowed ranges of $\tau$ and $n$ resulting in the positive valued $\omega$, to investigate
the stable solutions we find the allowed ranges of $\tau$ and $n$, given by Table 3 and Table 4, resulting in the negative values of ${\rm tr M}$ where use has been made of (\ref{Fbi60}) and (\ref{Fbi61}) with $c_1=c_2=0$. Note that, we should be careful that the following results are obtained in accordance with the consistency ranges of this model mentioned in Table 1 and Table 2. For instance, in the case $a(\tau)=\frac{1}{\tau^{n}}$ we have $-{\rm tr M}>0$ in the interval $\left(0.4<\tau<0.6~,~0<n<0.1\right)$; but in this range we face with the negative values of $\omega(\tau)$ and hence this interval
is ruled out.

 \vspace{5mm}
\begin{center}
{\scriptsize{ Table 3: }}\hspace{-2mm} {\scriptsize  Intervals for $\tau$ and $n$ resulting from the condition $-{\rm tr M}>0$ for the case $a(\tau)=\frac{1}{\tau^{n}}$.}\\
    \begin{tabular}{|l| l |l |  p{133mm} }
    \hline
   {\footnotesize$~~~~~~\tau$ }& ~~{\footnotesize~~~~~ $n$ }  \\ \hline
{\footnotesize ~$\left(0.01,0.37\right)$} & ~~{\footnotesize $\left[1,2.1\right)$}
\\\hline
 {\footnotesize ~$\left(0.42,+\infty\right)$} & {\footnotesize  ~$\left[1,50\right)$}
\\ \hline
{\footnotesize ~~~~$\left(1,7\right)$} & ~~{\footnotesize  $\left(0,0.23\right)$}
\\ \hline
{\footnotesize ~~~~$\left(3,7\right)$} & ~{\footnotesize  $\left(0.26,0.33\right)$}
\\ \hline

    \end{tabular}
\end{center}

 \vspace{5mm}
\begin{center}
{\scriptsize{ Table 4: }}\hspace{-2mm} {\scriptsize  Intervals for $\tau$ and $n$ resulting from the condition $-{\rm tr M}>0$ for the case $a(\tau)=e^{n\tau}$.}\\
    \begin{tabular}{|l| l |l |  p{133mm} }
    \hline
   {\footnotesize$~~~~~~\tau$ }& ~~{\footnotesize~~~~~ $n$ }  \\ \hline
{\footnotesize ~$\left(0,0.02\right)$} & ~{\footnotesize $\left(-0.28,-0.14\right)$}
\\\hline
 {\footnotesize ~~$\left(2.8,10\right)$} & ~~{\footnotesize  ~$\left(0.2,0.42\right)$}
\\ \hline
{\footnotesize ~~~$\left(0.8,1\right)$} & ~~{\footnotesize  ~~$\left(0.87,0.9\right)$}
\\ \hline
{\footnotesize~ $\left(0,0.35\right)$} & ~~{\footnotesize ~~ $\left(1,1.3\right)$}
\\ \hline
{\footnotesize ~$\left(0.68,0.73\right)$} & ~{\footnotesize ~ $\left(1.01,1.32\right)$}
\\ \hline
    \end{tabular}
\end{center}

It would be fruitful to survey the stability of the dark energy solutions (\ref{Fbi54}) and (\ref{Fbi55}) for the presented scale factors $a(\tau)=\frac{1}{\tau}$ and $a(\tau)=e^{n\tau}$, respectively. Obviously, the solution (\ref{Fbi57}) is not stable since the ranges of Table 4 do not cover it , thus, we do not have any dark energy solution for the case $a(\tau)=e^{n\tau}$. For the de Sitter solution described by (\ref{Fbi54}) , we should plot ${\rm tr M}$ by using (\ref{Fbi54}) for the limit ${\rm n}\rightarrow 1+0$ to find the negative values of ${\rm tr M}$ in the following intervals for $\tau$

\be\label{Fbi80}
{\rm for}~~n\rightarrow 1+0~~~~~~0.13<\tau<0.16~~~~~~{\rm and}~~~~~~1.13<\tau<1.4,
\ee
which is covered with the ranges of Table 3 and so the de Sitter solution of the case $a(\tau)=\frac{1}{\tau}$ is stable.

\section{Conclusions\label{Sec6}}
 In this paper, we have constructed non-minimal kinetic coupling bigravity models with two independent scalar fields. It has been shown that a wide range of expansion history of the universe can be explained by a solution of the bigravity model, particularly inflation or current accelerating expansion of the universe. This description is  different from the predictions of the models in massive gravity where the background metric is non-dynamical and
does not lead to the spatially flat homogeneous FRW cosmological solution. In the original scalar-tensor bigravity theory \cite{22} it has been shown that in spite of the difficulty of finding stable cosmological solutions , one can find some explicit stable solutions by studying the sign of the
trace of perturbation matrix extracted from the perturbed field equations. In this work, we have followed the mentioned approach by studying the diagrams of the eigenvalue equation condition (\ref{Fbi79}), which imposes a constraint on the trace of perturbation matrix ${\rm M}$ to take just negative values.
We considered the de Sitter universe  in the non-minimal kinetic coupling bigravity model and showed that the current model thoroughly endorses the de Sitter universe evolution as well as the scalar tensor bigravity with the choice of $a(\tau)=b(\tau)=c(\tau)$ for the solution $a(\tau)=\frac{1}{\tau}$, and then obtained the cosmological constant $\tilde{V}(\tau)=\tilde{V_{0}}=\Lambda$ with $\omega(\tau)=0$. For the accelerating solution $a(\tau)=e^{n\tau}$ we got the dark energy universe for the ranges obtained in the relation (\ref{Fbi57}).\\
Also, we  considered the constraint $-{\rm tr M}>0$ in equation (\ref{Fbi79}) and investigated the stability condition for the ranges of $\tau$ and $n$  mentioned in Table 3 and Table 4. It should be noticed that these intervals are satisfying the mentioned allowed ranges for  $\tau$ and $n$ in Table 1 and Table 2. In the first place, we focused on the results of Table 3 for the case $a(\tau)=\frac{1}{\tau^{n}}$. Seemingly, the first two couple ranges of $\tau$ and $n$ in Table 3 outline roughly the solution interval $\tau>0$ with $n>1$ reporting the stability of the quintessence universe as one of the solutions  offered in section 4. On the other hand, two latter couple ranges of $\tau$ and $n$ implying the interval $0<n<1$  in the early universe are interpreting the phantom universe. Moreover, we considered the de Sitter solution $a(\tau)=\frac{1}{\tau}$  and analyzed its stability in (\ref{Fbi80}) according to the final results reported in Table 3 which surely shows  that it is stable because the value $n=1$ is included in the ranges reported in
Table 3 . Furthermore, in the case of other suggested scale factor $e^{n\tau}$ with the corresponding values of $\tau$ and $n$ mentioned in  (\ref{Fbi57}), showing the dark energy solution, we  cannot distinguish any stability.
As a result, the solutions of non-minimal kinetic coupling bigravity model has some common stability intervals with that of \cite{34}.  Throughout the
paper, we just considered the homogeneous perturbation, which is independent of the spatial coordinates. This is because the inhomogeneous perturbation and/or anisotropic background  create ghost \cite{33}, and the superluminal mode \cite{34} in general leads to the violation of causality \cite{35}.

\appendix

\section{The calculations of Eqs.(\ref{Fbibi76}) and (\ref{Fbi79}) \label{AA}}

In the following, we extract equations Eqs.~(\ref{Fbibi76}) and (\ref{Fbi79}) first by using (\ref{Fbi32})
\be\label{Fbi70d}
L=K+\frac{\dot{K}}{K}-\frac{\dot{H}}{H}.
\ee

Having plugged the equations (\ref{Fbi32}) and (\ref{Fbi70d}) into the equations (\ref{Fbi62})-(\ref{Fbi65}) we can eliminate $L$ and $c$ as follows

\begin{align}\label{Fbi71d}
0=&2M_{g}^{2}\left(\dot{H}-H^{2}\right)+m^{2}M_{{\rm eff}}^{2}a(\tau)b(\tau)\left(1-\frac{K}{H}\right)+\epsilon \omega(\eta)\dot{\eta}^{2}
M_{g}^{2}+\nn&\frac{\kappa M_{g}^{2}}{a(\tau)^{2}}\left[\omega(\eta)\dot{\eta}^{2}\left(-6H^{2}+\dot{H}\right)+H\omega'(\eta)\dot{\eta}^{3}
+2H\omega(\eta)\dot{\eta}\ddot{\eta}\right],
\end{align}

\begin{align}\label{Fbi72d}
0=&2M_{g}^{2}\left(\dot{H}+2H^{2}\right)+m^{2}M_{{\rm eff}}^{2}\left(6a(\tau)^{2}-5a(\tau)b(\tau)-a(\tau)\frac{K b(\tau)}{H}\right)-2M_{g}^{2}a(\tau)^{2}\tilde{V}(\eta)+\nn&
\frac{\kappa M_{g}^{2}}{a(\tau)^{2}}\left[\left(\dot{H}+3H^{2}\right)\omega(\eta)\dot{\eta}^{2}+H\omega'(\eta)\dot{\eta}^{3}
+2H\omega(\eta)\dot{\eta}\ddot{\eta}\right],
\end{align}

\begin{align}\label{Fbi73d}
0=&2M_{f}^{2}\left(-\frac{K\dot{H}}{H}+K^{2}\right)+2m^{2}M_{{\rm eff}}^{2}\left(-\frac{a(\tau)^{3}K}{b(\tau)H}+\frac{a(\tau)^{3}K^{2}}{b(\tau)H^{2}}\right)-2\lambda\sigma(\zeta)\dot{\zeta}^{2}M_{f}^{2}+\nn&
\frac{2\beta M_{f}^{2}H^{2}}{b(\tau)^{2}K^{2}}\left[\sigma(\zeta)\dot{\zeta}^{2}\left(2\dot{K}+6K^{2}-3K\frac{\dot{H}}{H}\right)-K\sigma'(\zeta)\dot{\zeta}^{3}-
2K\sigma(\zeta)\dot{\zeta}\ddot{\zeta}\right],
\end{align}

\begin{align}\label{Fbi74d}
0=&2M_{f}^{2}\left(-2K^{2}-\frac{K\dot{H}}{H}\right)+m^{2}M_{{\rm eff}}^{2}\left(2\frac{K^{2}b(\tau)^{2}}{H^{2}}-\frac{a(\tau)^{3}K^{2}}{b(\tau)H^{2}}-\frac{a(\tau)^{3}K}{b(\tau)H}\right)+
2M_{f}^{2}\tilde{U}(\zeta)c(\tau)^{2}\nn&+\frac{\beta M_{f}^{2}}{c(\tau)^{2}}\left[\sigma(\zeta)\dot{\zeta}^{2}\left(-6K^{2}+3KL-\dot{K}\right)-K\sigma'(\zeta)\dot{\zeta}^{3}-
2K\sigma(\zeta)\dot{\zeta}\ddot{\zeta}\right].
\end{align}
By considering the equations  (\ref{Fbi71d}) and  (\ref{Fbi73d}), we find the following equation

\begin{align}\label{Fbi75d}
&H-\frac{m^{2}M_{{\rm eff}}^{2}}{2HM_{g}^{2}}a(\tau)b(\tau)\left(1-\frac{K}{H}\right)-\frac{\epsilon \omega(\eta)\dot{\eta}^{2}}{2H}
-\frac{\kappa }{2Ha(\tau)^{2}}\left[\omega(\eta)\dot{\eta}^{2}\left(-6H^{2}+\dot{H}\right)+H\omega'(\eta)\dot{\eta}^{3}
+2H\omega(\eta)\dot{\eta}\ddot{\eta}\right]=\nn&K+\frac{m^{2}M_{{\rm eff}}^{2}}{2M_{f}^{2}}\left(-1+\frac{K}{H}\right)\frac{a(\tau)^{3}}{b(\tau)H}-\frac{\lambda\sigma(\zeta)\dot{\zeta}^{2}}{2K}+
\frac{\beta H^{2}}{2b(\tau)^{2}K^{3}}\left[\sigma(\zeta)\dot{\zeta}^{2} \times \left(2\dot{K}+6K^{2}-3K\frac{\dot{H}}{H}\right)-
K\sigma'(\zeta)\dot{\zeta}^{3}-
2K\sigma(\zeta)\dot{\zeta}\ddot{\zeta}\right].
\end{align}
Having eliminated $\dot{H}$ from (\ref{Fbi71d}) and (\ref{Fbi72d}) and then from (\ref{Fbi73d}) and (\ref{Fbi74d}), we calculate

\be\label{Fbi76d}
\left(\omega(\eta)\dot{\eta}^{2}\left(\frac{\epsilon}{2}-\frac{9\kappa H^{2}}{2a(\tau)^{2}}\right)+a(\tau)^{2}\tilde{V}(\eta)\right)=3H^{2}+\frac{m^{2}M_{{\rm eff}}^{2}}{M_{g}^{2}}\left(3a(\tau)^{2}-3a(\tau)b(\tau)\right),
\ee

\be\label{Fbi77d}
\left(\sigma(\zeta)\dot{\zeta}^{2}\left(\frac{\lambda H^{2}}{2K^{2}}-\frac{9\beta H^{4}}{2b(\tau)^{2}K^{2}}\right)+b(\tau)^{2}\tilde{U}(\zeta)\right)=3H^{2}+\frac{m^{2}M_{{\rm eff}}^{2}}{M_{f}^{2}} \left(\frac{a(\tau)^{3}}{b(\tau)}-b(\tau)^{2}\right).
\ee
Combination of (\ref{Fbi75d}), (\ref{Fbi76d}) and (\ref{Fbi77d}) gives us the following equation

\begin{align}\label{Fbi78d}
0=&\left(K-H\right)-\frac{\tilde{U}(\zeta)b(\tau)^{2}K}{2H^{2}}+\frac{\tilde{V}(\eta)a(\tau)^{2}}{2H}-\frac{K}{4a(\tau)^{2}}\left[\omega(\eta)\dot{\eta}^{2} \left(\frac{H}{H}-
6H\right)+\omega'(\eta)\dot{\eta}^{3}+2\omega(\eta)\dot{\eta}\ddot{\eta}\right]+
m^{2}M_{{\rm eff}}^{2}\nn&\times\left[\frac{K}{2H^{2}M_{f}^{2}}\left(\frac{a(\tau)^{3}}{b(\tau)}-b(\tau)^{2}\right)-\frac{3}{2HM_{g}^{2}}
\left(\frac{a(\tau)^{2}}{b(\tau)}-b(\tau)^{2}\right)+\left(1-\frac{K}{H}\right)\left(\frac{a(\tau)^{3}H}{4b(\tau)H^{2}}-
\frac{a(\tau)b(\tau)}{4HM_{g}^{2}}\right)\right]-
\frac{\beta}{4b(\tau)^{2}}\nn&\times\left[\sigma\dot{\zeta}^{2}\left(\frac{2\dot{K}H^{2}}{K^{3}}+\frac{6H^{2}}{K}-\frac{3H\dot{H}}{K^{2}}\right) -\frac{\sigma'(\zeta)\dot{\zeta}^{3}H^{2}}{K^{2}}-\frac{2H^{2}\sigma(\zeta)\dot{\zeta}\ddot{\zeta}}{K^{2}}\right].
\end{align}
Regarding  the independent equations (\ref{Fbi72d}),  (\ref{Fbi76d}),  (\ref{Fbi77d}) and  (\ref{Fbi78d}) beside using  (\ref{Fbi62}), (\ref{Fbi63}),  (\ref{Fbi64}) and  (\ref{Fbi65}) help us to extract the following four perturbation equations

\begin{align}\label{Fbi79d}
0=&\delta\dot{H}\left(2+\frac{\omega(\eta)\kappa}{a(\tau)^{2}}\right)+\delta H\left(10H+\frac{2m^{2}a(\tau)^{2}}{H}+\frac{12\kappa H\omega(\eta)}{a(\tau)^{2}}-\frac{2\dot{H}}{H}-\frac{\omega(\eta)}{2H}-\frac{\kappa \omega(\eta) \dot{H}}{Ha(\tau)^{2}}\right)-\nn&\frac{2m^{2}a(\tau)^{2}}{H}\delta K+\delta f_{a}\left(12m^{2}a(\tau)^{2}-16H^{2}+\omega(\eta)\left(2-\frac{36\kappa H^{2}}{a(\tau)^{2}}+\frac{36\kappa H^{3}}{a(\tau)^{2}}\right)+4\dot{H}\right)-\nn&12m^{2}a(\tau)^{2}\delta f_{b}+\delta\dot{\eta}\left(\frac{\omega(\eta)\kappa \dot{H}}{a(\tau)^{2}}+\frac{12\omega(\eta)\kappa H^{2}}{a(\tau)^{2}}+2H^{2}-2\dot{H}-\frac{\omega(\eta)}{2}\right)+\omega(\eta)\delta \eta\nn&\times\left(-2H-\frac{\kappa \ddot{H}}{a(\tau)^{2}}+\frac{\kappa \dot{H}^{}}{Ha(\tau)^{2}}-\frac{12\kappa H\dot{H}}{a(\tau)^{2}}+\frac{36\kappa H^{3}}{a(\tau)^{2}}+\frac{\dot{H}}{2H}\right)+\delta\eta\left(-14H\dot{H}+16H^{3}-2\ddot{H}+\frac{2\dot{H}^{2}}{H}\right),
\end{align}

\begin{align}\label{Fbi80d}
0=&\delta f_{a}\left(6H^{2}+\omega(\eta)\left(\frac{18\kappa H^{2}}{a(\tau)^{2}}-\frac{1}{2}\right)-6m^{2}a(\tau)^{2}\right)+\delta f_{b}\left(6m^{2}a(\tau)^{2}\right)+\delta H\left(-\frac{9\kappa H\omega(\eta)}{a(\tau)^{2}}-6H\right)+\delta\dot{\eta}\left(\frac{1}{2}-\frac{9\kappa H^{2}}{a(\tau)^{2}}\right)\nn&\times\omega(\eta)+\delta\eta \left(\omega(\eta)\left(\frac{9\kappa H\dot{H}}{a(\tau)^{2}}-\frac{18\kappa H^{3}}{a(\tau)^{2}}+\frac{H}{2}\right)+6H\dot{H}-6H^{3}\right),
\end{align}

\begin{align}\label{Fbi81d}
0=&\delta f_{a}\left(-6m^{2}a(\tau)^{2}\right)+\delta f_{b}\left(6H^{2}-\frac{\omega(\eta)}{2}+\frac{18\kappa \omega(\eta)H^{2}}{a(\tau)^{2}}+6m^{2}a(\tau)^{2}\right)+\delta\dot{\zeta}\times\left(\frac{\omega(\eta)}{2}-\frac{9\kappa \omega(\eta)H^{2}}{a(\tau)^{2}}\right)+\nn&\delta H\left(\frac{\omega(\eta)}{2H}-\frac{18\kappa \omega(\eta)H}{a(\tau)^{2}}-6H\right)+\delta K\left(-\frac{\omega(\eta)}{2H}+\frac{9\kappa \omega(\eta)H}{a(\tau)^{2}}\right)+\delta \zeta \left(\omega(\eta)\left(\frac{9\kappa H\dot{H}}{a(\tau)^{2}}-\frac{18\kappa H^{3}}{a(\tau)^{2}}+\frac{H}{2}\right)+6H\dot{H}-6H^{3}\right),
\end{align}

\begin{align}\label{Fbi82d}
0=&\left(\delta K-\delta H\right)\left(\omega(\eta)\left(-\frac{15\kappa }{4a(\tau)^{2}}+\frac{\kappa \dot{H}}{2a(\tau)^{2}H^{2}}+\frac{3}{8H^{2}}\right)-\frac{3}{2}+\frac{\dot{H}}{H^{2}}\right)\nn&+\left(\delta f_{a}-\delta f_{b}\right)\left(4H-\frac{\dot{H}}{H}+\omega(\eta)\left(\frac{9\kappa H}{2a(\tau)^{2}}-\frac{1}{2H}\right)\right)+\left(\delta \omega(\eta)-\delta \sigma(\zeta)\right)\frac{9\kappa H}{4a(\tau)^{2}}\nn&+\left(\delta\eta-\delta\zeta\right)\left(\omega(\eta)\left(\frac{3\kappa \dot{H}}{a(\tau)^{2}}-\frac{9\kappa H^{2}}{a(\tau)^{2}}+\frac{1}{2}-\frac{\dot{H}}{8H^{2}}+\frac{\kappa \ddot{H}}{4a(\tau)^{2}H}-\frac{\kappa \dot{H}^{2}}{4a(\tau)^{2}H^{2}}\right)+\frac{\ddot{H}}{2H}-\frac{\dot{H}^{2}}{2H^{2}}+\frac{7\dot{H}}{2}-4H^{2}\right)+\nn&\left(\delta \dot{\eta}-\delta \dot{\zeta}\right)\left(\omega(\eta)\left(\frac{1}{8H}+\frac{3\kappa H}{2a(\tau)^{2}}-\frac{\kappa \dot{H}}{4a(\tau)^{2}H}\right)+\frac{\dot{H}}{2H}-\frac{H}{2}\right)+\left(\delta \dot{H}-\delta \dot{K}\right)\frac{\kappa \omega(\eta)}{2a(\tau)^{2}H}.
\end{align}
Apparently, the equation (\ref{Fbi80d}) determines $\delta \dot{\eta}$ by which we can extract $\delta \dot{H}$ from (\ref{Fbi79d}) as
\be\label{Fbi83d}
\delta \dot{\eta}=\frac{-b_{1}}{b_{4}}\delta f_{a}-\frac{b_{2}}{b_{4}}\delta f_{b}-\frac{b_{3}}{b_{4}}\delta H-\frac{b_{5}}{b_{4}}\delta\eta.
\ee

\be\label{Fbi84d}
\delta \dot{H}=\delta H\left(-\frac{h_{2}}{h_{1}}+\frac{h_{6}b_{3}}{h_{1}b_{4}}\right)+\delta K\left(-\frac{h_{3}}{h_{1}}\right)+\delta f_{a}\left(-\frac{h_{4}}{h_{1}}+\frac{h_{6}b_{1}}{h_{1}b_{4}}\right)+\delta f_{b}\left(-\frac{h_{5}}{h_{1}}+\frac{h_{6}b_{2}}{h_{1}b_{4}}\right)+\delta\eta \left(-\frac{h_{7}}{h_{1}}+\frac{h_{6}b_{5}}{h_{1}b_{4}}\right).
\ee
Equation (\ref{Fbi81d}) gives us $\delta \dot{\zeta}$ as
\be\label{Fbi85d}
\delta \dot{\zeta}=-\frac{c_{1}}{c_{3}}\delta f_{a}-\frac{c_{2}}{c_{3}}\delta f_{b}-\frac{c_{4}}{c_{3}}\delta H -\frac{c_{5}}{c_{3}}\delta K-\frac{c_{6}}{c_{3}}\delta\zeta.
\ee
Finally, in order to obtain $\delta\dot{ K}$, we substitute the equations  (\ref{Fbi83d})- (\ref{Fbi85d}) into (\ref{Fbi82d}) as follows

\be\label{Fbi86d}
\delta \dot{ K}=g_{1}\delta H+g_{2} \delta K+g_{3}\delta f_{b}+g_{4}\delta f_{a}+g_{5}\delta\eta+g_{6}\delta\zeta,
\ee
where
\begin{align}\label{Fbi87d}
&g_{1}=-\frac{h_{2}}{h_{1}}+\frac{h_{6}b_{3}}{h_{1}b_{4}}-\frac{b_{3}d_{4}}{b_{4}d_{5}}+\frac{c_{4}d_{4}}{c_{3}d_{5}}-\frac{d_{1}}{d_{5}},\nn&
g_{2}=-\frac{h_{3}}{h_{1}}+\frac{c_{5}d_{4}}{c_{3}d_{5}}+\frac{d_{1}}{d_{5}},\nn&
g_{3}=-\frac{h_{5}}{h_{1}}+\frac{h_{6}b_{2}}{h_{1}b_{4}}-\frac{b_{2}d_{4}}{b_{4}d_{5}}+\frac{c_{2}d_{4}}{c_{3}d_{5}}-\frac{d_{2}}{d_{5}},\nn&
g_{4}=-\frac{h_{4}}{h_{1}}+\frac{h_{6}b_{1}}{h_{1}b_{4}}-\frac{b_{1}d_{4}}{b_{4}d_{5}}+\frac{c_{1}d_{4}}{c_{3}d_{5}}+\frac{d_{2}}{d_{5}},\nn&
g_{5}=-\frac{h_{7}}{h_{1}}+\frac{h_{6}b_{5}}{h_{1}b_{4}}-\frac{b_{5}d_{4}}{b_{4}d_{5}}+\frac{d_{3}}{d_{5}},\nn&
g_{6}=\frac{c_{6}d_{4}}{c_{3}d_{5}}-\frac{d_{3}}{d_{5}},
\end{align}

and also

\begin{align}\label{Fbi88d}
&h_{1}=2+\frac{\kappa\omega}{a^{2}},\nn&
h_{2}=10H+\frac{2m^{2}a^{2}}{H}+\frac{12\kappa H\omega}{a^{2}}-\frac{2\dot{H}}{H}-\frac{\omega}{2H}-\frac{\kappa \omega \dot{H}}{a^{2}H},\nn&
h_{3}=-\frac{2m^{2}a^{2}}{H},\nn&
h_{4}=12m^{2}a^{2}-16H^{2}+\omega\left(2-\frac{36\kappa H^{2}}{a^{2}}\right)+4\dot{H},\nn&
h_{5}=-12m^{2}a^{2},\nn&
h_{6}=\omega\left(\frac{\kappa \dot{H}}{a^{2}}+\frac{12\kappa H^{2}}{a^{2}}-\frac{1}{2}\right)+2H^{2}-2\dot{H},
\end{align}

\begin{align}\label{Fbi89d}
&b_{1}=6H^{2}+\omega\left(\frac{18\kappa H^{2}}{a^{2}}-\frac{1}{2}\right)-6m^{2}a^{2},\nn&
b_{2}=6m^{2}a^{2},\nn&
b_{3}=-\frac{9\kappa\omega H}{a^{2}}-6H,\nn&
b_{4}=\omega\left(\frac{1}{2}-\frac{9\kappa H^{2}}{a^{2}}\right),\nn&
b_{5}=\omega\left(\frac{H}{2}+\frac{9\kappa H\dot{H}}{a^{2}}-\frac{18\kappa H^{3}}{a^{2}}\right),
\end{align}

\begin{align}\label{Fbi90}
&c_{1}=-b_{2},\nn&
c_{2}=-b_{1},\nn&
c_{3}=\omega\left(\frac{1}{2}-\frac{9\kappa H^{2}}{a^{2}}\right),\nn&
c_{4}=\omega\left(\frac{1}{2H}-\frac{18\kappa H}{a^{2}}\right)-6H,\nn&
c_{5}=\omega\left(\frac{-1}{2H}+\frac{9\kappa H}{a^{2}}\right),\nn&
c_{6}=b_{5},
\end{align}

\begin{align}\label{Fbi91d}
&d_{1}=\omega\left(\frac{-15\kappa}{4a^{2}}+\frac{\kappa \dot{H}}{2a^{2}H^{2}}+\frac{3}{8H^{2}}\right)-\frac{3}{2}+\frac{\dot{H}}{H^{2}},\nn&
d_{2}=4H-\frac{\dot{H}}{H}+\omega\left(\frac{9\kappa H}{2a^{2}}-\frac{1}{2H}\right),\nn&
d_{3}=\omega\left(\frac{3\kappa \dot{H}}{a^{2}}\frac{-9\kappa H^{2}}{a^{2}}+\frac{1}{2}-\frac{\dot{H}}{8H^{2}}+\frac{\kappa \ddot{H}}{4a^{2}H}-\frac{\kappa \dot{H}^{2}}{4a^{2}H^{2}}\right)+\nn&\frac{\ddot{H}}{2H}-\frac{\dot{H}^{2}}{2H^{2}}+\frac{7\dot{H}}{2}-4H^{2}+
\frac{9\kappa H}{4a^{2}}F(a),\nn&
d_{4}=\omega\left(\frac{1}{8H}-\frac{3\kappa H}{2a^{2}}-\frac{\kappa \dot{H}}{4a^{2}H}\right)+\frac{\dot{H}}{2H}-\frac{H}{2},
\end{align}
where $\delta\sigma(\zeta)=F(a)\delta\zeta$,  $\delta\omega(\eta)=F(a)\delta\eta$ and  $F(a)$ is of the order  $a^{3}$ or $H^{3}$.
Herein, looking back to the equations (\ref{Fbi46}) and (\ref{Fbi48}) shows that we cannot work with the analytical exact expression for $\omega(\tau)$ and $\sigma (\tau)$ unless we specify the scale factor $a(\tau)$ by which we become able to solve (\ref{Fbi46}).
Any way, using the equations (\ref{Fbi83d})-(\ref{Fbi86d}) beside the two following equations 

\be\label{Fbi93d}
\delta H=\delta\dot{f_{a}}~,~~~~~\delta K=\delta\dot{f_{b}}~,
\ee
we seek the range of stability of the solutions.
It is worthwhile to note that $\omega(\tau)$ appears in the final form of the relations (\ref{Fbi83d}), (\ref{Fbi84d}), (\ref{Fbi85d}), (\ref{Fbi86d})
and (\ref{Fbi93d}) in which we replace it  by  (\ref{Fbi60}) and  (\ref{Fbi61}) (it should be remarked that we have chosen $c_{1}=c_{2}=0$) for $a(\tau)=\frac{1}{\tau^{n}}$ and $a(\tau)=e^{n\tau}$, respectively.

\section{The expressions for ${\rm\textbf{ {tr M}}}$  \label{AB}}

 For the ${\rm tr M}$, we have

\begin{align}\label{ap1}
{\rm tr M}=&\left(24 a^{10}\left(\omega-4 H^{2}+2\dot{H}\right)+\right.\nn
& \left.72 \omega^{3} H^{2}\left(-477 H^{4}+\dot{H}+3 H^{2}\left(4+27 \dot{H}\right)\right)+
2 a^{8}\left(1488 H^{4}+\omega\left(13 \omega+74 \dot{H}\right)- 4 H^{2}\left(169 \omega+258\dot{H} \right)\right)+\right.\nn
& \left.6\omega^{2}a^{2}H^{2}\left(-6\omega-11052H^{4}+\left(8-57\omega\right)\dot{H}+3H^{2}\left(136+223\omega+924\dot{H}\right)\right)
+\omega a^{4}\times
\right.\nn& \left.
\left(-26928H^{6}+5\omega^{2}\dot{H}+96H^{4}\left(16+187\omega+114\dot{H}\right)-H^{2}\left(\omega\left(96+157\omega\right)+
36\left(8+79\omega\right)\dot{H}\right)\right)+2a^{6}H^{2}\times
\right.\nn& \left.
\left(-1440H^{4}+\frac{\omega^{2}}{H^{2}}\left(\omega+36\dot{H}\right)+
124H^{2}\left(8+839\omega+84\dot{H}\right)-\left(\omega\left(24+613\omega\right)+
\frac{12\dot{H}}{H^{2}}\left(8+223\omega\right)\right)\right)\right)\times\nn&
\left(2H\omega\left(\omega+2a^{2}\right)\left(a^{2}-18H^{2}\right)\left(12a^{4}-36H^{2}\omega+a^{2}\left(\omega-12H^{2}\right)\right)\right)^{-1}.
\end{align}

\end{document}